# A Conversation with Pranab Kumar Sen

**Malay Ghosh and Michael J. Schell**

*Abstract.* Pranab Kumar Sen was born on November 7, 1937 in Calcutta, India. His father died when Pranab was 10 years old, so his mother raised the family of seven children. Given his superior performance on an exam, Pranab nearly went into medical school, but did not because he was underage. He received a B.Sc. degree in 1955 and an M.Sc. degree in 1957 in statistics from Calcutta University, topping the class both times. Dr. Sen's dissertation on order statistics and nonparametrics, under the direction of Professor Hari Kinkar Nandi, was completed in 1961. After teaching for three years at Calcutta University, 1961–1964, Professor Sen came to Berkeley as a Visiting Assistant Professor in 1964. In 1965, he joined the Departments of Statistics and Biostatistics at the University of North Carolina at Chapel Hill, where he has remained.

Professor Sen's pioneering contributions have touched nearly every area of statistics. He is the first person who, in joint collaboration with Professor S. K. Chatterjee, developed multivariate rank tests as well as time-sequential nonparametric methods. He is also the first person who carried out in-depth research in sequential nonparametrics culminating in his now famous Wiley book *Sequential Nonparametrics: Invariance Principles and Statistical Inference* and SIAM monograph. Professor Sen has over 600 research publications. In addition, he has authored or co-authored 11 books and monographs, and has edited or co-edited 11 more volumes. He has supervised over 80 Ph.D. students, many of whom have achieved distinction both nationally and internationally. Professor Sen is the founding co-editor of two international journals: *Sequential Analysis* and *Statistics and Decisions*. He is a Fellow of the American Statistical Association and of the Institute of Mathematical Statistics, and an elected member of the International Statistical Institute. Professor Sen was the third recipient of the prestigious Senior Noether Award offered by the Nonparametrics Section of the American Statistical Association. In 2007, a Festschrift was held in his honor at the Nonparametrics Conference on the 70th anniversary of his birth.

This conversation took place at the Speech Communication Center, University of North Carolina at Chapel Hill on November 11, 2005.

*Malay Ghosh is Distinguished Professor, University of Florida, Department of Statistics, P.O. Box 118545, Gainesville, Florida 32611-8545, USA e-mail: ghoshm@stat.ufl.edu. Michael J. Schell is Professor and Biostatistics Department Chair, Moffitt Cancer Center, MRC/BIOSTAT, 12902 Magnolia Drive, Tampa, Florida, USA e-mail: michael.schell@moffitt.org.*

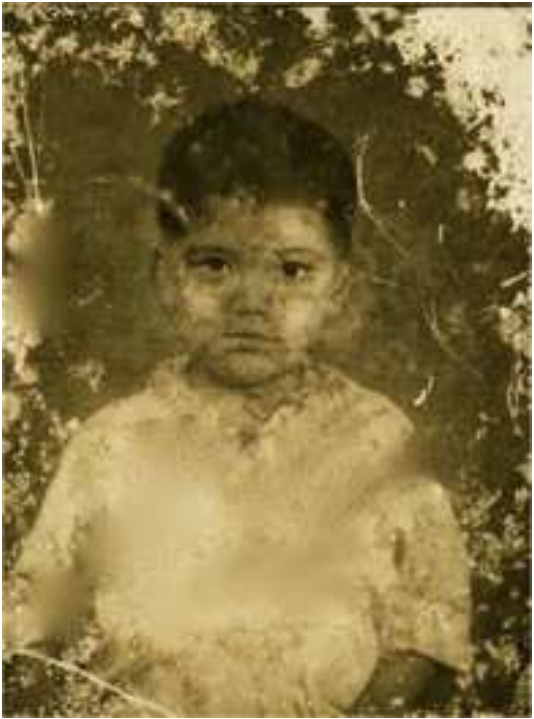

Fig. 1. *Pranab Sen at age 5.*

## EARLY YEARS AND COLLEGE DAYS

**Schell:** Good morning, Dr. Sen. To begin our conversation, tell us a bit about your early childhood years.

**Sen:** Well, I really appreciate this occasion to converse with both of you, Malay and Michael. The whole event of life is full of unforeseen and unaccountable happenings starting with birth, propagating all the way until someone closes the eyes forever. In my case, I was born in a not too affluent but educated family; my father was a railway officer; my mother was the daughter of a noted physician in herbal medicine, and the pre-secondary school days went out pretty smoothly until I was about 10 years old when my 43-year-old father died of leukemia, the first detected case in India. That was the first significant stochastic event in my life and it continued to have a deleterious impact for years. My mother, gifted with enormous patience, was thrust with the responsibility of raising seven children. I was the second one, and much later I could appreciate how diligently she handled the whole matter. When I was in high school I almost gave up my studies; I was very restless and more involved with sports and other distractive social events. Gradually, my mother led me through those difficult days until I entered my final school year. In my school days, I never had a top standing, until in the tenth grade, against all expectations, I topped the school list in the annual as well as in the matriculation examinations; indeed a chancy event beyond my expectation, the only deterministic factor being my mother's strong pursuance.

**Ghosh:** You talk about your life in poetry as one filled with chancy events (your verse "My Chancy Life as a Statistician" reproduced in "*Pranab Kumar Sen*: *Life and Works*" by Balakrishnan, Pena and Silvapulle, 2008); please tell us about some that led you to become a statistician.

**Sen:** When I was in the high school, my mother, based on her (Ayurvedic physician) father's earnest desire, was hoping all the time that I could go to medical school. With that intention, I was admitted to the Intermediate Science Section in R. G. Kar Medical College in Calcutta; the principal, realizing my financial difficulties, told me that if I could secure the top score among all students in my class, I would be given a full 5-year scholarship to study medicine. I was delighted with that challenging offer which made me more serious in my studies. Another chancy event—when the examinations were over and I was about to apply formally and confidently for admission to the medical college, they realized that I was 18 months underage. I was advised to pursue a two-year B.Sc. degree and then return to the medical program; the scholarship based on my earlier examination performance would remain intact. But my uncle advised me to not take that risk because if he died in the middle of my anticipated and circumstantially enhanced 7 years medical studies, I would have to quit (without a formal medical degree) and take some clerical job to support my brothers and sisters who were still in school. So I had to withdraw myself from that possibility. I applied for admission and was selected into both the Mathematics and Physics honors programs in the Presidency College, Calcutta. Another casting of a die: Prabir Acharya, a school friend of mine, came to see me then and suggested that I not take any one of them but rather negotiate for Statistics honors where I could do even better. I convinced my mother and uncle and switched to Statistics honors in the same college; that brought me to statistics in the first place which I wasn't at all planning.

**Ghosh:** What did you think about statistics, once you entered the program?

**Sen:** I started appreciating more and more this novel discipline during my undergraduate years in the Presidency College and subsequently at the Calcutta University, located just across the street from the Presidency College. We had wonderful and most dedicated teachers in both places. Professor Anil Bhattacharya, a well-known statistician (famous for the Bhattacharya information bound and divergence measure), was our teacher at the Presidency College. Professors B. N. Ghosh and P. K. Banerjee taught us at both the Presidency College and the University, while Professors P. K. Bose, M. N. Ghosh, H. K. Nandi, K. N. Bhattacharya, K. C. Seal and others taught us at the University. They had tremendous insight and had a profound impact on my perceptions, interest and career development in statistics. My uncle urged me not to pursue the Masters degree after completing the B.Sc. honors program, but to take a job to support my brothers' and sisters' educations (as in my elder brother's case). But, as I did well in my B.Sc. honors program, I was given the signal to go for two more years in the M.Sc. program. There too I did well, but my family's financial situation led me to seek a job. I went to the Indian Statistical Institute, Calcutta (ISI), hoping for an assistantship that could enable me to pursue a doctoral degree while providing financial support to my family. I was interviewed and aptitude tested at ISI but was not selected. I was advised to take a regular administrative job in the Demography (Vital Statistics) Department with the Government of West Bengal. I realized, though, that although that job might well satisfy my financial needs, it would not lead to my career objectives. Thus, I came back to my alma mater, Calcutta University Statistics Center and was most heartily welcomed by all my teachers there. Because of the usual irregularity of receipt of stipend, I took some extra tutorial jobs to support my siblings' educations. Professors Manindra Nath Ghosh and Hari Kinkar Nandi were the two teachers who supported me most in this venture. Professor Ghosh left for another academic position in the Institute of Agricultural Research Statistics, New Delhi, just before our M.Sc. examinations were over, but he kept a vigilant eye on me and continued interacting when needed.

I was really fortunate to have Professor Hari Kinkar Nandi as my advisor, guru and mentor for four years which I cherished most deeply; I could visualize a broader interpretation of the subject, beyond the fenced domain of mathematical statistics and probability theory, and it broadened my interest in such a profound way that, even after earning a Ph.D. degree, I continued to have an appreciation of statistics encompassing its interdisciplinary applications as well as statistical theory.

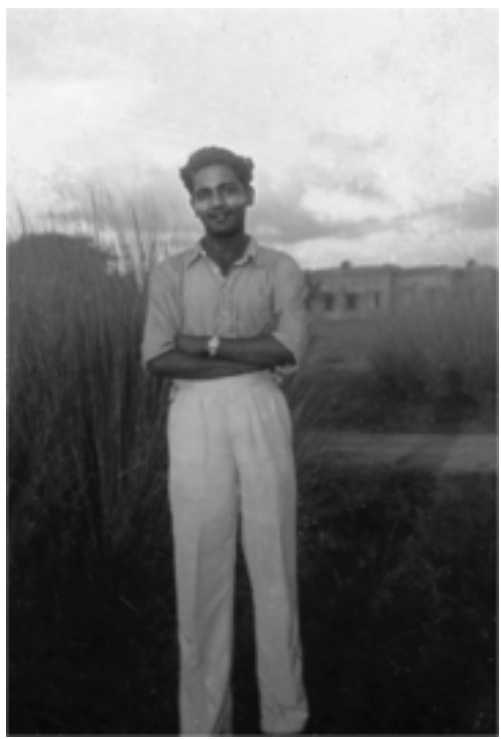

Fig. 2. *While in Presidency College, Calcutta, 1954.*

**Ghosh:** Interesting. Obviously Professor Hari Kinkar Nandi had a clear impact on you, and many of us were his students also; can you tell us more about him, his mentorship of you, and his style?

**Sen:** He did not hold a formal Ph.D. degree in Statistics or an allied subject but was the top mentor for about 20 doctoral students, many of whom have done extremely well. He was the founding editor of the *Calcutta Statistical Association Bulletin*, and with dedication and distinction, he edited the journal for about 30 years. Although I was formally his first advisee, at least four others before me worked under his supervision, albeit unofficially. I was fortunate to have strong friendships with two fellow students, Shoutir Kishore Chatterjee and Jayanta Kumar Ghosh. We used to sit in a small room with barely the leg space for three small desks and nothing else. Library and computing facilities were inadequate, but not our morale; our strength was our mentor Professor Nandi and the unique environment created within the walls of those small rooms constituting the first Statistics department in India. We were given a complete free hand to choose our own dissertation topics but with helpful hands whenever we needed. I could appreciate how that helped us develop the spirit of appraising contemporary research with a view to exploring further work.

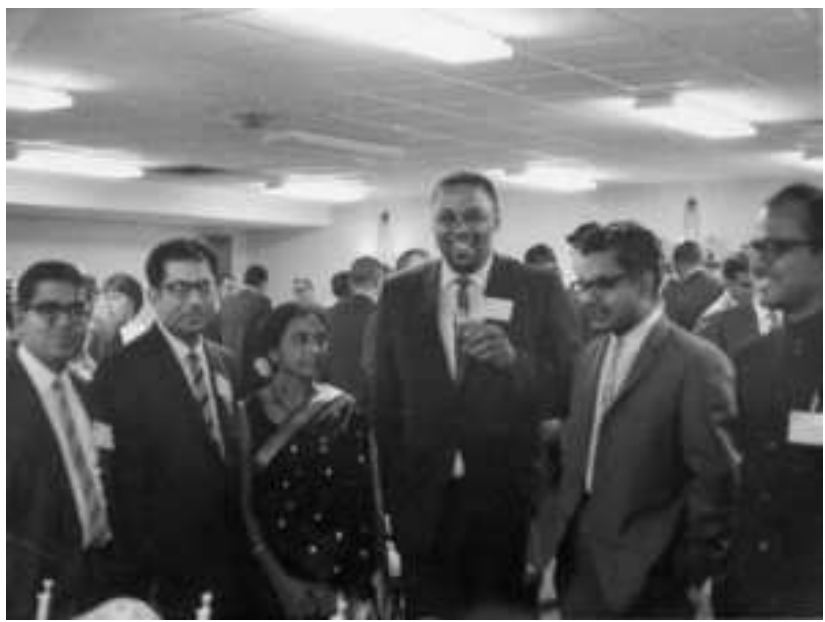

Fig. 3. *Jayaram Sethuraman, Madan and Uma Puri, Chuck Bell, Pranab Sen and Shoutir Chatterjee at the first Nonparametrics Conference, Bloomington, IN, 1969.*

Professor Nandi was a bachelor. In his undergraduate class, he did very well in Physics Honours, and he moved to Statistics at the Masters level. Design and analysis of experiments, multivariate analysis, sample survey and statistical inference (including decision theory) were his primary areas of interest, although he had profound knowledge in many other fields. What caused me the most wonder was that he

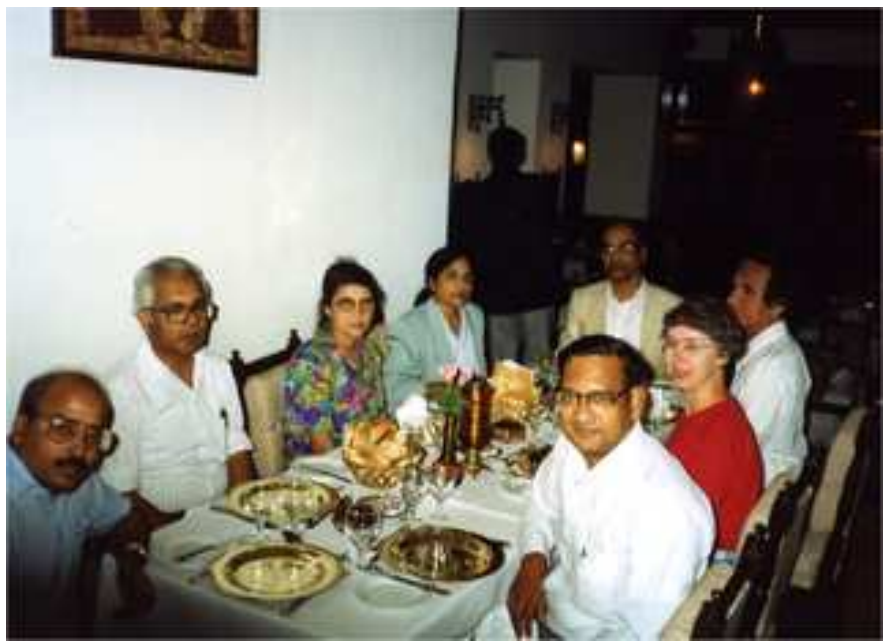

Fig. 4. *Bikas Sinha, Pranab and Gauri Sen, Mrs. and Dr. A. K. Md. E Saleh, J. K. Ghosh and Mrs. and Dr. A. P. Basu around a restaurant table in Cairo, 1991.*

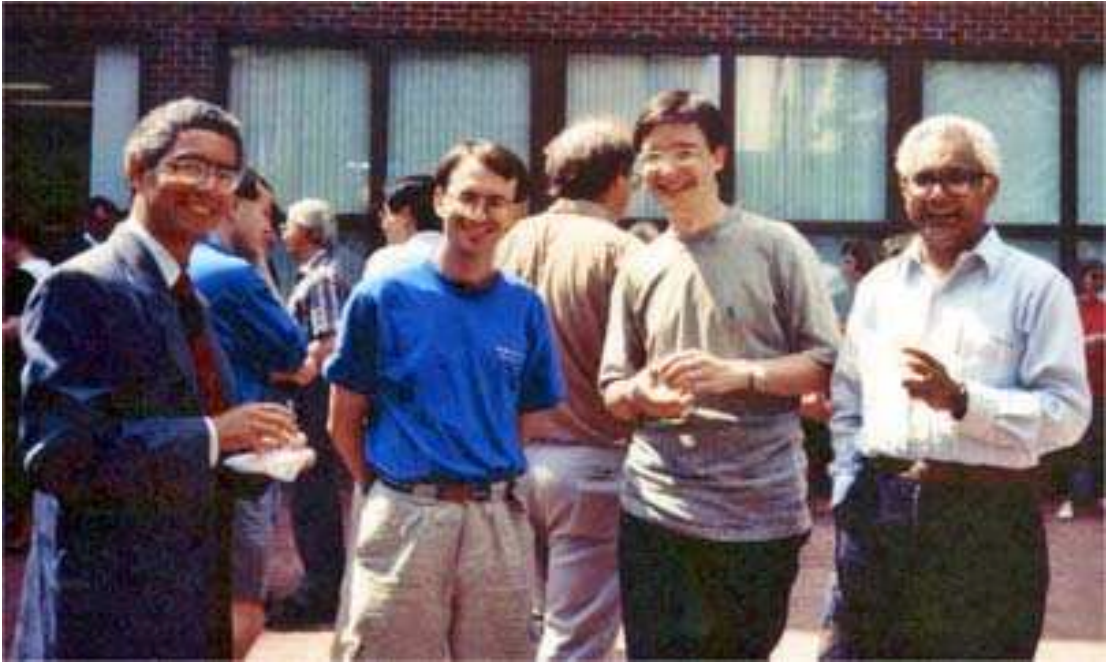

Fig. 5. *Julio Singer, John Preisser, Antonio Pedroso de Lima and Pranab Sen in the McGavran–Greenberg Building courtyard, 1994.*

suggested diverse areas of research to his advisees so that we would each feel comfortable working in our area of specializations. This trend continued with his later advisees too, including S. R. Chakraborti,

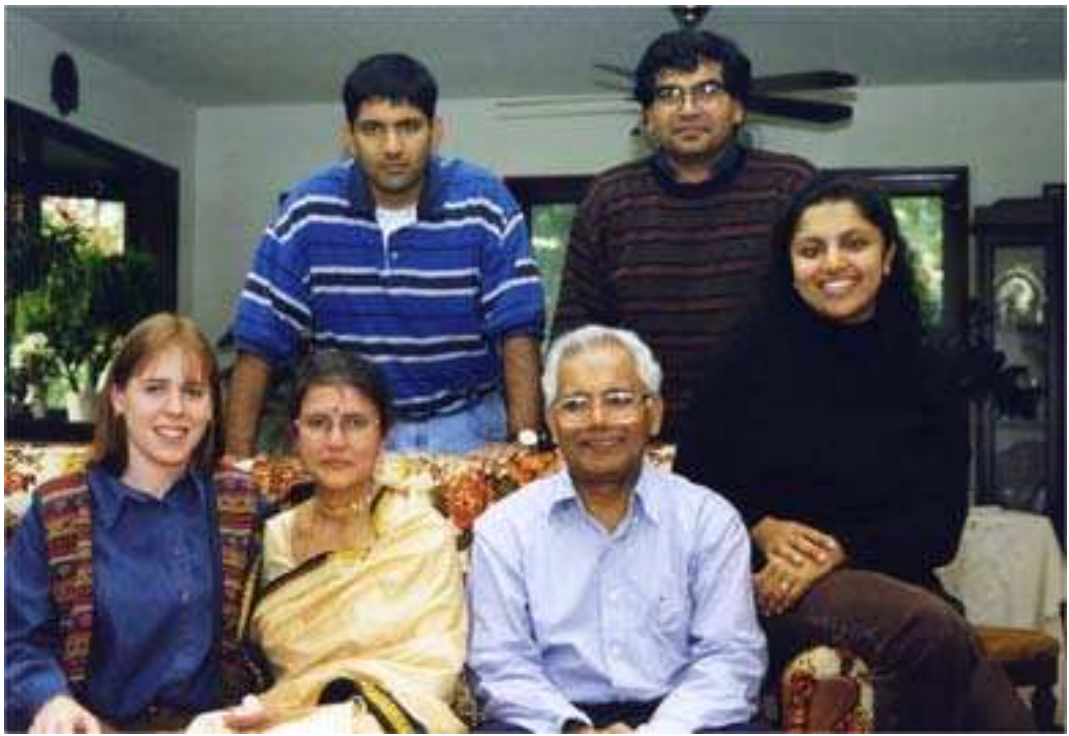

Fig. 6. *Kim, Gauri, Pranab, Devi, Ru and Joy (standing) at home in Chapel Hill, 1997.*

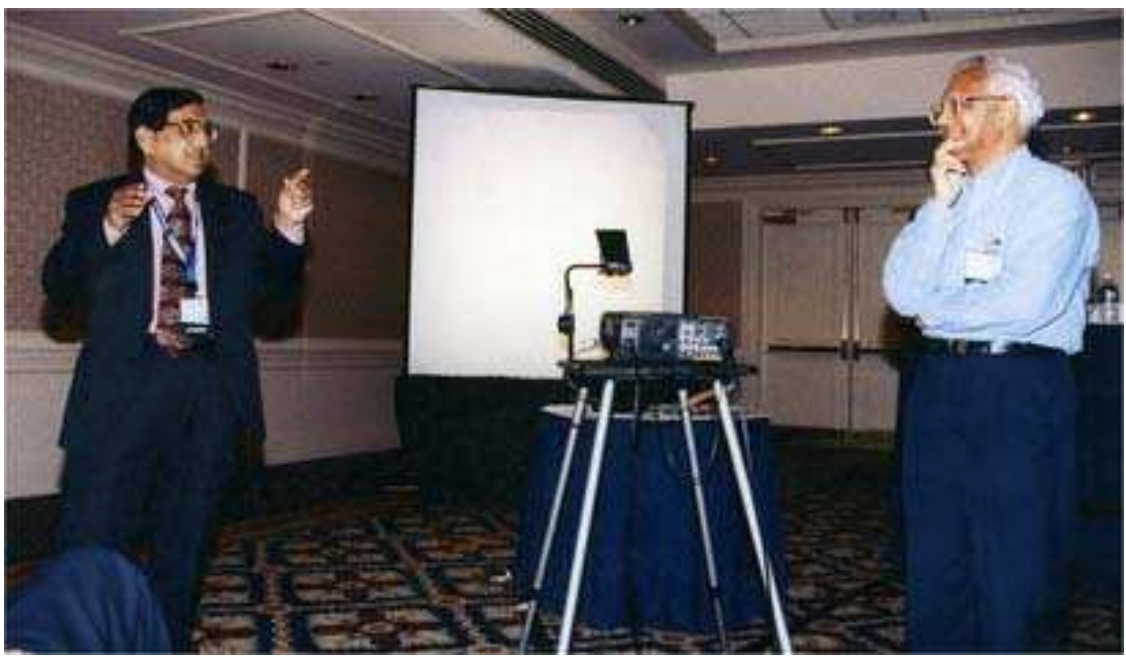

Fig. 7. *Malay Ghosh raises a question for Pranab Sen as the 2002 Noether Award speaker, Joint Statistical Meetings, New York City.*

Basudeb Adhikari, Arijit Chaudhuri and Bimal and Bikas Sinha (the statistical twins). A few years after joining Calcutta University, he started studying homeopathic and biochemical medicine on his own. We used to wonder how detached a person could be and still how devoted was he to the advancement of postgraduate training and research in statistics in the Indian subcontinent. It was indeed a golden opportunity for us to appreciate his wisdom, patience and mentorship.

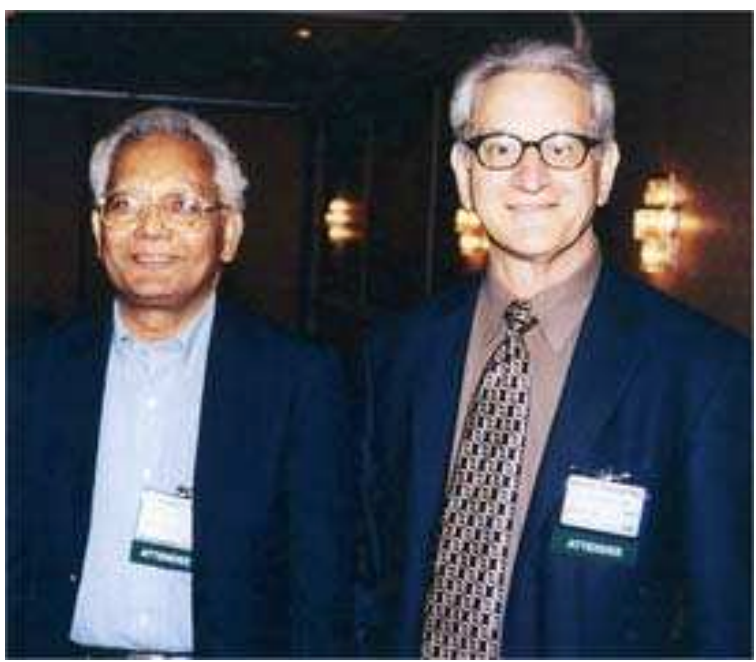

Fig. 8. *Pranab Sen and Ralph D'Agostino at the Noether Lecture, Joint Statistical Meetings, 2002.*

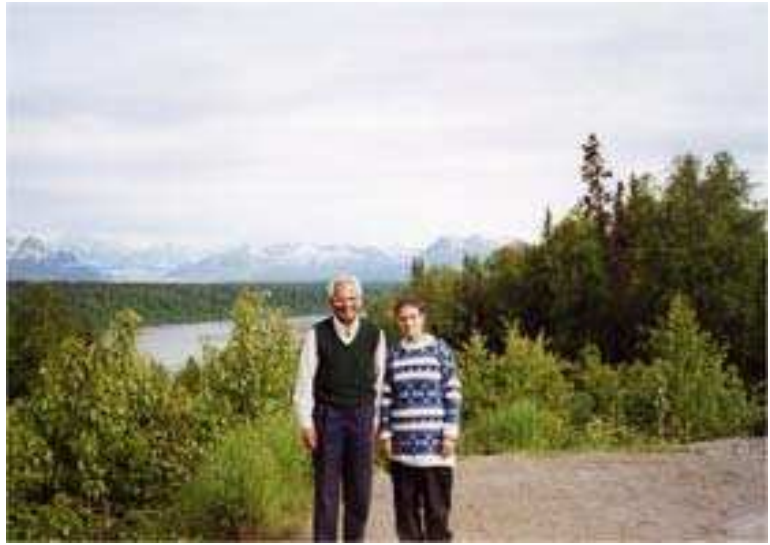

Fig. 9. *Gauri and Pranab in Alaska, 2002.*

## RESEARCH AT CALCUTTA UNIVERSITY

**Ghosh:** Tell us a bit about your research at Calcutta University.

**Sen:** As I started teaching in Calcutta University in 1961, I was asked to teach a course on Biological Assay. I realized at that time that full attention was being paid to only the normal tolerance distribution. I thought that that was very unrealistic because, in most cases, tolerance distributions were very skewed, and even after suitable dosage transformation, were not near to normal ones. My next sister, Malaya, in her dissertation work in physiology, had a lot of data from Calcutta Medical School on triglycerides and other blood chemicals; their distributions being heavily skewed. I observed that even after log transformations, near-normality was not achieved. This inspired me to develop nonparametric methods for biological assays. The 1963 *Biometrics* paper (Sen, 1963a) was probably the very first nonparametric one in bioassay. I observed that because ranks are invariant under arbitrary strictly monotone transformations, an estimator of the rela-tive potency based on the ranks (viz. the Wilcoxon–Mann–Whitney statistic) would enjoy the same property; that is, it would not matter whether we worked with the log of the dose, or any other transformation. This invariance eliminates all the arbitrariness of dosages that underlies the use of the conventional normal tolerance distribution.

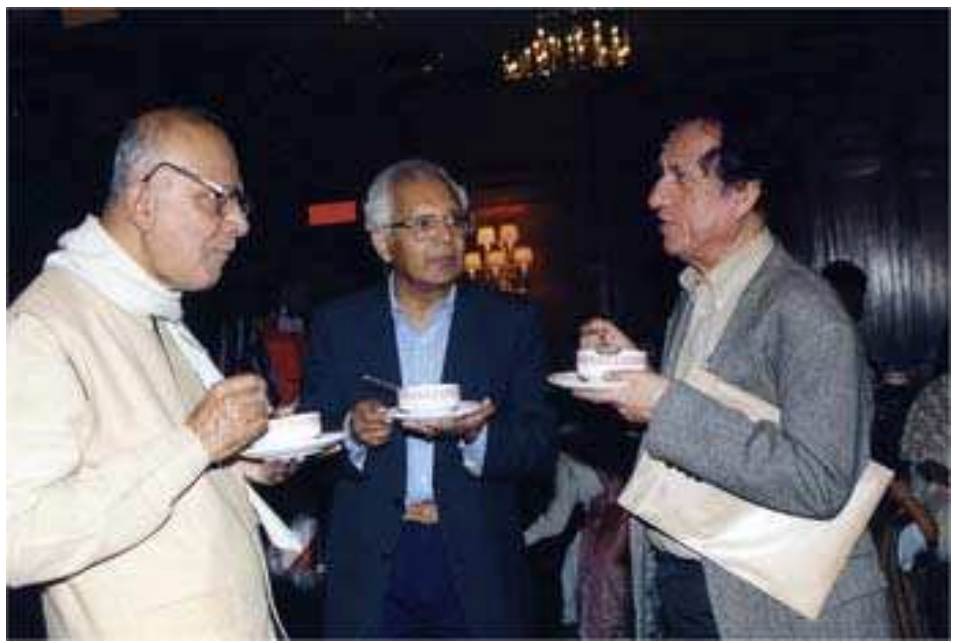

Fig. 10. *Shoutir Chatterjee, Pranab Sen and Jayanta Ghosh, Kolkata, India, 2003.*

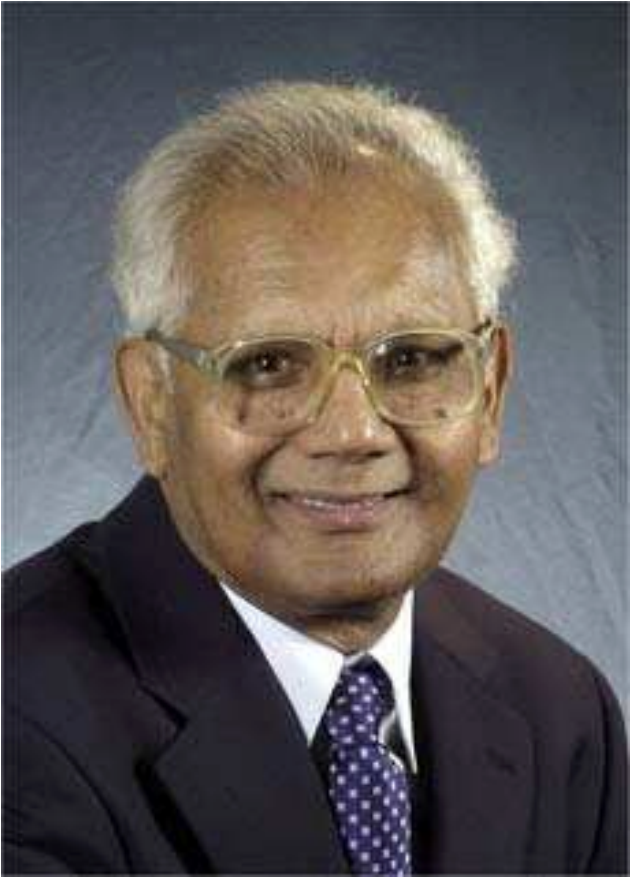

Fig. 11. *Pranab Sen at his office in Chapel Hill, 2007.*

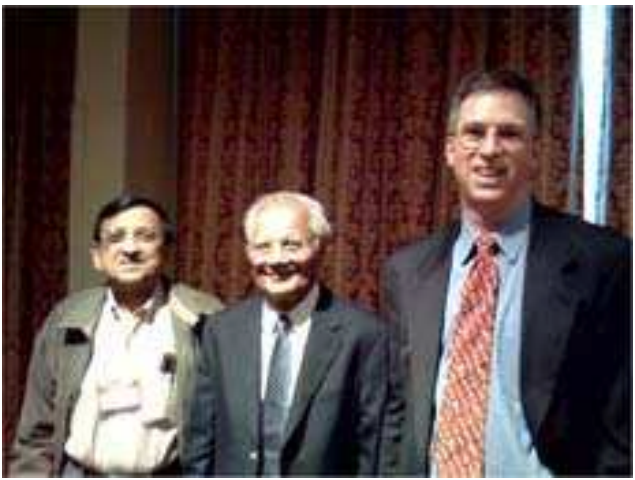

Fig. 12. *Malay Ghosh, Pranab Sen and Michael Schell at the 70th Birthday Festschrift for Pranab Kumar Sen, annual Nonparametrics Conference, Columbia, SC, 2007.*

After completing my Ph.D. work in order statistics and nonparametrics, I was looking for new frontiers of research. I converged with Shoutir Chatterjee to a common domain: multivariate analysis from his expertise and nonparametrics from mine! The whole area of multivariate nonparametric analysis flourished in Calcutta in 1963 (Sen, 1963b); and our first paper (Chatterjee and Sen, 1964) on bivariate two-sample location problems, some 40 pages long, provided all the impetus to probe into this fertile area in subsequent years.

I was also able to extend the findings of my 1960 U-statistics paper (Sen, 1960) to dependent sequences arising in time series models, and with the help of Professor Nandi to finite population sampling (Nandi and Sen, 1963), the latter being the very first work on the theory of U-statistics in finite population sampling. Even now I see some people using such results in complex sampling schemes, survival analysis and time series analysis.

**Schell:** In 1964 you left India to take a visiting professorship at Berkeley. Was that a difficult step to take?

**Sen:** It was less difficult in 1964 than two years earlier when I was offered exactly the same position but could not come partly due to family responsibilities. By 1964 my sister Malaya had earned her Ph.D. degree in physiology and brother Mander completed his engineering degree. A shower of chances: I got three offers within a span of a week: one from the Lehigh University in Bethlehem, Pennsylvania, the second from the University of California, Berkeley, and the third from University of North Carolina, Chapel Hill. The UNC offer was largely due to M. N. Ghosh's effort; actually, I learned later on that he was offered a senior position at UNC but could not come, and he recommended me strongly to Dr. Greenberg. I recall that Professor S. N. Roy, the guru of our gurus and a most respected professor at UNC, while visiting Calcutta in the fall of 1962, became aware of my offer from Berkeley, and also wanted me to come to Chapel Hill to work with him. All forces united behind the Chapel Hill offer in 1964. However, the week before the UNC offer,

I was offered a one-year visiting position by Professor Lucien LeCam from Berkeley, which I accepted promptly. I informed Professor Greenberg of my acceptance of the Berkeley offer, and he replied: "No problem. Go to Berkeley and we will snatch you from there next year." And that is exactly how I came to Chapel Hill in 1965. Was it not a chancy outcome?

## MOVE TO CHAPEL HILL

**Schell:** You came to Chapel Hill in the fall of 1965, which is 40 years ago, and you obviously have loved it here because you are still here at UNC, so the random walk of your life seemed to slow down a bit.

**Sen:** It is in a sense true but even after coming here, I had not decided whether I should stay for a long time or should go back to Calcutta. I remember during the three years (1964–1967) when I was on leave of absence from Calcutta University, I used to write my both affiliations on all my publications. Some of my Chapel Hill colleagues used to ask me whether I was serious about continuing this dual affiliation! I had to defend myself—Calcutta University was my home and I couldn't give it up. Eventually, I realized that UNC was one of the best places for statistics in America, if not the world, and by being here I could not only strengthen my background but develop additional ties with Indian schools. This way UNC induced me to settle in Chapel Hill, despite several offers from other universities over the years. Before 1964, my attractions for UNC were mostly due to Professors Hotelling, Hoeffding, Johnson and Roy in the Statistics Department and Greenberg in Biostatistics. Since Professor S. N. Roy passed away in 1964 before my arrival in Chapel Hill, I realized that I could revive my research interest on order statistics here with Johnson, Greenberg and H. A. David, who came a year earlier. My aspiration was to strengthen my mathematics background and at the same time dip into the high-tides of applications which would make the statistical research relevant.

**Schell:** Tell me about some of your early research efforts.

**Sen:** As I said earlier, it was very nice of Professor Nandi to let us choose our own dissertation topics. In the course of this search, I came across a paper of Hotelling and Chu on the moments of the sample median (Chu and Hotelling, 1955). I could see that their finding could be extended to a general class of sample quantiles; that led to a 1959 paper of mine (Sen, 1959) which constituted the first part of my dissertation. I also came across a classical paper on the theory of U-statistics by Professor Wassily Hoeffding. I extended the results and wrote a paper which occupied another major part of my dissertation work and was published in 1960 (Sen, 1960). This paper contained all the basics of a variance estimation technique now known as the jackknifing variance estimator. Dr. W. J. (Jack) Hall, who was also at Chapel Hill at that time, brought it to the attention of other researchers. Professor Hoeffding got interested in this work and wrote a UNC technical report in 1961 (Hoeffding, 1961), where he mentioned that Sen had considered the result under a moment condition of order greater than 1 while he wanted to prove the result assuming only the first moment. He used an ingenious martingale method, but got stuck at some point and was so absorbed in other things that he left it unfinished. In the meanwhile, Bob Berk (1966) considered a reversed martingale approach in a different context, yielding as a byproduct the almost sure convergence of U-statistics under the first moment. This martingale theory for U-statistics reshaped the study of asymptotics not only for U-statistics but also for general nonparametric statistics. Most of these developments as well as extensions of my 1963 bioassay paper (Sen, 1963a) took place after I came to Chapel Hill, and that left me with a deep sense of satisfaction. All these convinced me to make Chapel Hill my second home; actually, my wife and I have spent more time here than anywhere else!

## FAMILY

**Schell:** Very interesting. Your family flourished when you came to Chapel Hill; tell us more about your wife, Gauri, and your children, Devadutta and Aniruddha.

**Sen:** We got married in 1963. Gauri came from a more affluent family; her father was an outstanding man in many respects, a science graduate with an interest in management. At that time I worried how her family background could match with mine. Now, without hesitation, I could say that she was very adaptive. My birth family lived with all my uncles and aunts together, some 30 people in a household. Her parents and my mother arranged for the negotiated marriage; I first saw her on our wedding day. My mother was so happy to see this young daughter-in-law from a very different family background adapt well in our household. After coming to Chapel Hill, I could see that if I stayed, I would need the maximum support from her—and it still continues—but how wonderful she has been all these years, not only supporting me but fulfilling the dreams of my mother.

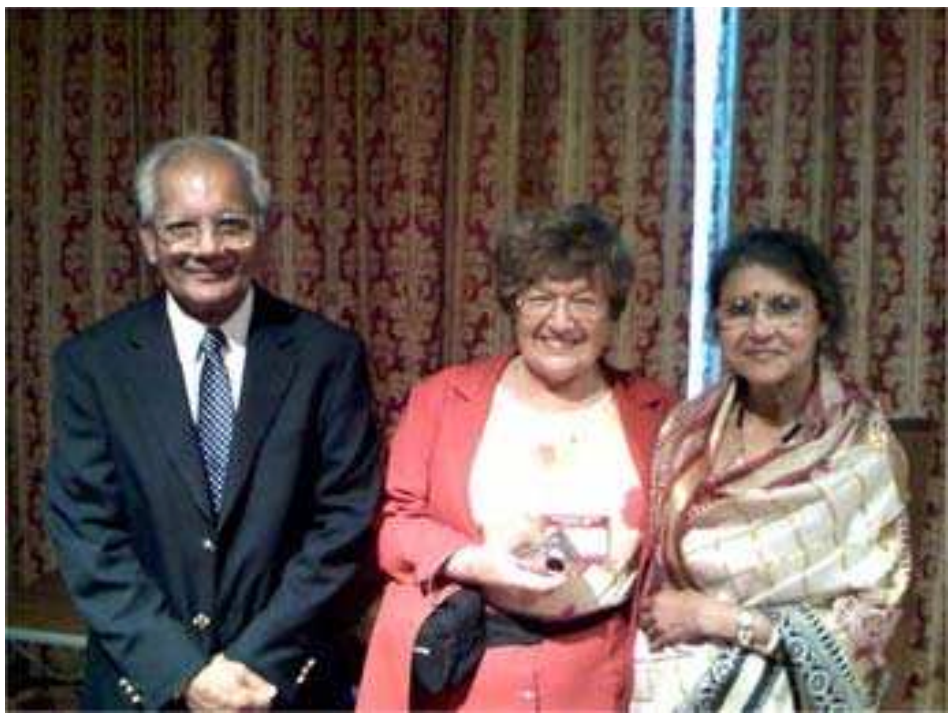

Fig. 13. *Pranab Sen, Jana Jurečková and Gauri Sen at the 70th Anniversary Festschrift for Pranab Kumar Sen, Columbia, SC, 2007.*

Our two children, Devadutta (daughter) and Aniruddha (son), were born in Chapel Hill in October 1966 and February 1973, respectively. Devi majored in Journalism at UNC, and worked for newspapers in Fort Myers, Florida, and Danville, Illinois, and Atlanta (the Constitution). She is presently pursuing creative writing at home while raising her three active children. Our son, called Ru by his friends, majored in Speech Communication at UNC. During his studies he became very strongly religiously minded and now he is a minister in a Presbyterian church here. He has a son, Jacob Kalyan, bearing our family name, and a daughter, Lily. We are very happy with both our children and all our grandchildren, too.

## MAJOR CAREER ACOMPLISHMENTS

**Ghosh:** We talked about your career working on all the statistics and nonparametrics. From a citation perspective, your peak years occurred in the years 1966–1975. Articles that you wrote during those 10 years have received more than 1,000 citations. Can you tell us about your work in these years?

**Sen:** You know, Malay, I am not that citation minded and have never paid any attention to citations because I have the conviction that creativity and citations are two different attributes; combining the two needs a special talent, and I confess that I am no good in that perspective. The current fashion of counting citations often overlooks the early developments. Is it not true that the citation business flourished in the 1980s, and hence publications prior to 1970 have often been put in the backburner! Just to emphasize this point let me mention the ISI citation in mathematical sciences covering the period 1991–2000; you will see that out of more than 200 top citations, although some 80 are from the statistics area, they did not include the pioneers whose fundamental contributions in the 1950s and 1960s reshaped the statistics discipline. Indeed citation scoring is a different kind of game.

**Ghosh:** I know that you don't care about citations, but I still like to draw attention to one of your nonparametric papers, Estimates of the Regression Coefficient based on Kendall's Tau (Sen, 1968). This paper has received nearly 250 citations (386 as of March 2009). The interesting thing is the procedure that you describe has now become known as the Sen–Theil Approach. What made you think about this procedure?

**Sen:** Well, that paper was generated from my 1963 *Biometrics* paper (Sen, 1963a) which dealt with the two-sample problem; I was convinced that something true for the two-sample case must be capable of yielding a similar picture for the simple regression problem. When I looked at Kendall's Tau and its invariance under any strictly monotone transformation on the observations, I said to myself: "This is wonderful, because all the entities in this statistic are the signs of the divided difference of pairs of observations and their regressors!" Therefore, the same structure exploited in the Wilcoxon two-sample case can be used here. It appeared that the sample median of these divided differences, yielding the point estimate of the slope, reduces to the estimate of the shift in the two-sample case. Further, the distribution-free confidence interval in the two-sample case obtained in 1963 goes over smoothly for the simple regression model. At the same time, this simple characterization does not hold in general for estimates based on general linear rank statistics, even in the two-sample model. There is a need for iterative solutions; that was the major initiative.

James Adichie, in 1967, considered estimation of the regression slope based on linear rank statistics (Adichie, 1967). Although his work was very interesting, he could not come up with an explicit form of the point estimator nor of a distribution-free confidence interval. H. Theil, an eminent econometrician, had run into similar problems but got some very interesting and motivating results, especially for the case where the regressors are all distinct. Though not a mathematical statistician, Theil really provided a clear-cut direction long before my 1963 *Biometrics* paper or the contemporary work of Hodges and Lehmann (1963). Establishing the invariance and other statistical properties (including asymptotic normality, consistency, and possible extension to measurement error models), I was able to go one more step in statistical interpretation and application. I could mention also that my 1968 paper led to some other work by Jana Jurečková (1969), Hira Koul (1969) and you and me too where some refined results on "asymptotic linearity of rank statistics in regression parameters" were developed and exploited in the study of asymptotic properties of estimators (Ghosh and Sen, 1971). Coming back to your point on citations, I have the feeling if the citation web were in effect from 1960s, instead of 1980s, the picture could have been different!

**Ghosh:** Let me come back to your 1963 *Biometrics* paper. It is true that you have independently proposed a nonparametric estimator of relative potency that was essentially the same as the rank estimator of shift parameter considered by Hodges and Lehmann (1963) at the same time, and is commonly referred to as the Hodges–Lehmann estimator. Should not we term it as the Hodges–Lehmann–Sen estimator?

**Sen:** Again, Malay, I won't go for any profound claim, but if you and others who know the field well want to do so, I won't raise any serious objection. In his 1974 *Nonparametrics* book (Lehmann, 1974), Lehmann pointed out somewhere that for bioassay models, Sen (1963a) considered the same estimator. To bypass this accreditation problem, some people simply refer to that as the Wilcoxon score R-estimator. It may not be out of the way to mention that while you and others may debate on this small point, in 1966 Professor Noether pointed out to us a 1952 book chapter (in Walker and Lev) by Lincoln Moses (Moses, 1953), who explored the median of midranges of all possible distinct pairs, known as the Walsh median, just a bit different from the Wilcoxon score estimator, and in 1965, Dr. Moses provided a graphical display of the Wilcoxon score two-sample point as well as confidence interval estimator (Moses, 1965). Thus, perhaps we should all share the cake of creditation and resolve any potential controversies amiably! As a matter of fact, in the 1967 Joint Statistical Meetings in Washington D.C., Gottfried Noether organized an invited paper session with Erich Lehmann and me, and him as the third speaker. As Erich could not come, Peter Bickel was his substitute. I talked about nonparametric confidence intervals, exploiting my 1966 *Annals* paper (Sen, 1966), and going beyond to the Kendall Tau statistic-based ones. Noether was so interested, knowing that you were about to earn your Ph.D. degree on related topics, he invited you to his department and offered you a faculty position.

**Ghosh:** That's correct. I remember that I presented there some work on a sequential confidence interval based on the Kendall Tau statistic, and it was very well received. I know that your area of expertise covers parametrics, nonparametrics as well as semiparametrics which were yet to be popular in the early 1970s. Those of us around in Chapel Hill at that time used to regard you as a mathematical statistician, albeit in residence at a biostatistics department. We knew that you had genuine interest in applications, and yet up to that time, your work was predominantly in theory and methodology, with occasional detours in probability theory and stochastic processes. Could you explain what inspired your transition to more application-oriented methodology research?

**Sen:** Malay, I would like to iterate that the broad training in statistics I had in India led to my conviction not to limit myself to narrow sub-areas of mathematical statistics. My inclination to continue at UNC was primarily influenced by the setup of the Statistics Department in the College of Arts and Science and Biostatistics in the School of Public Health, within the Division of Health Affairs. I was given a free hand to work in both departments and thereby could appreciate the tremendous scope of statistics in the mathematical as well as clinical and public health disciplines. In fact, Dr. Greenberg was fully aware of my background and aspiration, and made it clear that it would be highly beneficial for the Biostatistics Department if I continued working with good methodological problems while my colleagues collaborated on health and clinical research applications. This was the best opportunity I could expect at any place. Professors Jim Grizzle, Robert Elston, Gary Koch and others were more intensely devoted to applied work, and I could collaborate with them whenever they had any methodological problems and there was no dearth of that. Jim Grizzle was involved with messy data problems in medical studies, and had many practical insights which needed methodological support, and this provided the orientation of my growing interest in applied work. I recall that Malay, Jim and I worked out some nonparametrics for growth curve analysis in the early 1970s which appeared in the *Journal of the American Statistical Association* (Ghosh, Grizzle and Sen, 1973).

**Ghosh:** Please describe the LIPIDS project for which you developed some interesting methods.

**Sen:** In the fall of 1971, the UNC Biostatistics and Epidemiology Departments were awarded a long-term clinical coordination task by the National Heart, Lung and Blood Institute, called the LIPIDS project. Eight hospitals across the country were coordinated for a multicenter longitudinal study of the impact of lowering blood cholesterol level on reduction of cardiovascular disease risk. Some 3,952 healthy males between the ages of 35 and 60 years, with cholesterol levels of 230 or more, were included in a double-blind study with two groups: placebo and treatment, of almost equal number. Their failure patterns were to be statistically appraised. The NIH team had a very plausible hypothesis mingled with medical ethics and operational cost constraints. The study could be continued to a maximum of 12 years' period (July 1972 to June 1984), but if at any intermediate point of time, the null hypothesis of no difference between the placebo and treatment group was rejected in favor of the treatment group, the trial should be stopped with the surviving subjects switched to the treatment group; thus the null hypothesis would only be accepted if there was no compelling statistical evidence of better survival for the treatment group throughout the study. Since this study involved a common cohort of subjects, the time-sequential failures were neither independent nor identically distributed. On top of that there were many explanatory variables so that underlying failure distributions were not simple. The classical Wald sequential probability ratio test was deemed inapplicable, and there was right-truncation due to the imposed twelve-year duration as well as possible dropout and noncompliance. I was asked to develop the statistical methodology for this sequential procedure, with interim analyses to be appraised every 3 months. Fortunately Professor Shoutir Chatterjee from Calcutta was visiting UNC in 1972, and together we dipped into the nature of the stochastic processes arising in such schemes, without assuming any specific parametric models. Another chancy event: We observed that under the null hypothesis, a general class of linear rank statistics has a simple martingale structure that can be incorporated in a permutation setup for nonparametric analysis and yet can be attuned to the Wald sequential probability ratio test theory by transforming calendar time to information time. This long paper (Chatterjee and Sen, 1973) provided the access to suitable applications in the LIPIDS as well as other clinical trials.

Malay, you may recall that maybe a year or two later, we worked on martingale properties of conventional rank statistics with respect to the sample size variation. These two related research questions absorbed me completely for a decade (1972–1982).

**Ghosh:** Linking nonparametrics with sequential methods with application to clinical trials?

**Sen:** Yes, with many advisees, this work culminated in *Sequential Nonparametrics* (Sen, 1981b). The most appealing point in this effort was the basic feature that statistical methodology can really open up a chain of fruitful applications in clinical trials, time-sequential procedures, repeated significance tests and survival analysis. Most probably, due to this work, I was endowed with a distinguished professorship at UNC in January 1982 when I was 44 years old; also I was designated as NSF/CBMS Lecturer in Statistics, and presented a set of 10 lectures on "Theory and Application of Sequential Nonparametrics" at the University of Iowa. This area of research is now flooded with statisticians, going beyond parametrics, and it gives me great satisfaction to know that back in the 1970s the few of us who were pursuing methodological research in applied clinical problems were able to provide the impetus for others to join the camp.

**Schell:** This important development occurred about the same time as the classical work of D. R. Cox (1972). Could you comment on the impact of his work on yours?

**Sen:** This seminal work of David Cox was undoubtedly a masterpiece; it developed a proportional hazard model which laid down the foundation of semiparametrics. The beauty of this paper is the motivation and general formulation. The mathematical foundations were developed later by Cox (1975) and more rigorously by many others in the late 1970s and early 1980s going over to the so-called multiplicative intensity processes. With my own inclination on martingale characterizations of various statistics in 1981, I was also able to characterize the general asymptotics for the Cox model, through martingales for induced order statistics or concomitants of order statistics (Sen, 1981a). The proportional hazards assumption, basic to the Cox model, needs to be critically appraised in any real application. Jim Grizzle and I came across a case with the congestion in the upper aorta for elderly people where some statisticians were blindly using a proportional hazard model for the surgery and medication groups. However, the hazard functions were quite different and crisscrossed; hence, Cox model-based analysis was not ideal. There are countless such instances, and I hope that semiparametricians who love the Kaplan–Meier estimator (Kaplan and Meier, 1958) and the Cox proportional hazard model for their mathematical convenience would check the appropriateness of such simplifying assumptions in their specific contexts. I had a conversation with David Cox quite some time ago, and he also echoed similar sentiments. I wish that more of us would have such insight.

**Schell:** You have also done tremendous work on sequential analysis and on a mixture of nonparametric and sequential methods. When did you first get interested in sequential analysis?

**Sen:** In 1958–1962, Shoutir Chatterjee and Jayanta Ghosh were both working on sequential methods, while I focused on nonparametrics. Interactions with them initiated my interest in sequential methods as well, albeit with an eye on linking it to nonparametrics. However, my active involvement dawned about a decade later when Malay was a postdoc at UNC, and we launched on nonparametric sequential methods in a systematic way using martingale theory. This kept me busy during 1971–1983. Back in 1972, to this sequential arena, with active collaboration with Chatterjee, we were able to annex the time-sequential analysis based on nonparametrics.

**Schell:** One of your biggest strengths is the use of sophisticated asymptotic theory in virtually every area of statistics. How did you develop the skill?

**Sen:** The thrust for asymptotics started with my dissertation work back in 1958–1959. The main inspiration came from Professor M. N. Ghosh; his asymptotic work in the early 1950s on serial rank statistics drew my interest but I realized that I would need to strengthen my mathematical background. The UNC Statistics complex provided me with the golden opportunity to further this objective and again I would also acknowledge the tremendous inspiration and support I received from Dr. Greenberg in Biostatistics. I extended my work on functional central limit theorems and invariance principles for U-statistics as well as rank statistics with such novel asymptotics. Back in 1960, the level of asymptotics was set by the 1948 seminal work of Hoeffding on U-statistics (Hoeffding, 1948) and the more recent work of Chernoff and Savage (1958) on rank statistics. The contiguity-based approach in Hájek (1962) set another direction to this asymptotics while the Pyke–Shorack (Pyke and Shorack, 1968a, 1968b) and Hájek (1968) work led to additional avenues. I was fortunate to be abreast of these developments, and using martingale methods, we extended the contiguity approach to a more general asymptotic setup. These developments culminated in my 1981 sequential nonparametric book (Sen, 1981b), but are still very much relevant to my methodological work in some applied problems.

**Ghosh:** In addition to your nearly 600 publications, you have authored or co-authored 11 books and monographs, and edited or co-edited 11 more. How did you find time for these things?

**Sen:** Malay, I am in the academics for about 45 years, and these developments did not take place overnight! I had a modest number of publications prior to moving to UNC, but collaboration with a large number of colleagues and advisees resulted in this spectrum, so my credit has to be discounted accordingly! From 1965–1971, Dr. Greenberg enabled me to devote 75% time to research, and that really helped in completing the 1971 book on multivariate nonparametrics with Madan Puri. During the 1970s, I had little time to write monographs. However, beginning again in the 1980s, I had more opportunity to complete *Sequential Nonparametrics* (1981, 1985) (Sen, 1981b), *Handbook of Statistics, Volume 4: Nonparametric Methods* with Dr. Krishnaiah (Krishnaiah and Sen, 1984), *Nonparametric Methods in General Linear Models* with Madan Puri (Puri and Sen, 1985), and a couple of *Festschrifts* in honor of Norman Johnson and Bernard Greenberg. The 1990s were more devoted to book writing: *Large Sample Methods in Statistics* with Julio Singer (Sen and Singer, 1993); *Pitman's Measure of Closeness* with Jerry Keating and Bob Mason (Keating, Mason and Sen, 1993); *Robust Statistical Procedures* with Jana Jurećková (Jurećková and Sen, 1996); and *Sequential Estimation* with Nitis and you (Ghosh, Muknopadhyay and Sen, 1997).

Back in 1994, Mrs. Hájek and Professor Šidák asked me to undertake a thorough revision of the classic Hájek–Šidák *Theory of Rank Tests* book (Hájek and Šidák, 1967). After much effort, the second, enlarged edition came out from Academic Press in 1999 (Hájek, Šidák and Sen, 1999). Back in the 1980s I was interested in constrained statistical inference, especially in nonparametric setups. Two year-long visits of Mervyn Silvapulle ultimately resulted in the 2004 book on this broad area encompassing parametrics as well as beyond parametrics (Silvapulle and Sen, 2004).

**Ghosh:** Let's talk about your famous book with Madan Puri on multivariate rank tests published in 1971 (Puri and Sen, 1971). What was the intended audience for this book?

**Sen:** Lehmann's 1959 hypothesis testing book (Lehmann, 1959) and the Hájek–Šidák *Theory of Rank Tests* set the audience at the statistics graduate level. This led us to consider a little less abstract treatise of the subject matter, yet still aimed at the same level. We used to share a joke at that time: Hájek–Šidák (1967) and Puri–Sen (1971) led Erich Lehmann to write an even simpler nonparametrics book in 1974. If I would have rewritten this 1971 text, it would have been more in line with our 1997 sequential estimation level, emphasizing statistical interpretations more than asymptotics.

**Ghosh:** You co-edited two volumes for the *Handbook of Statistics* and contributed numerous articles to the *Encyclopedia of Statistical Science* (and Biostatistics). What impact do you think that these publications had?

**Sen:** In the 1970s, I was caught in the middle of abstract theoretical developments and the much needed applications where the methodology would be very helpful. Faced with this dilemma, the two *Handbook of Statistics* volumes, nonparametrics with P. R. Krishnaiah (Krishnaiah and Sen, 1984), and bioenvironmental and public health with C. R. Rao (Sen and Rao, 2000), were an effort to illustrate the role of statistical methodology in various interdisciplinary fields of applications. The encyclopedia articles were of the nontechnical expository type for the convenience of users who lack more complete statistical expertise; it has served well from a reference material perspective, and I am happy to see that such methods are getting more attention in recent applied works.

**Ghosh:** You are an editor of *Sequential Analysis*. Can you comment on the stature of that journal?

**Sen:** The thrust for specialized journals arose in the 1970s, partly fueled by the shortage of space in the society journals (AMS, JASA) and partly to accommodate more in-depth presentation of specialized work. I was on the founding editorial board of the *Journal of Multivariate Analysis* (1972) and the *Communications in Statistics*, around the same time. My college friend Bhaskar Ghosh (at Lehigh) was very much in the mainstream of sequential hypothesis testing, writing a well-received book around 1970, while time-sequential, repeated significance testing, group sequential methods, and nonparametrics were mostly developed later in that decade. So, in 1980 when Marcel Dekker requested that we start a new journal in this area, we discussed the project with a number of active researchers in this field. I am happy to say that we received overwhelming support and many of them joined the editorial board. After 15 years, we handed over the editorial responsibilities to Malay, who dutifully promoted the area for an additional 8 years. The present editorial board is much more diversified with many more members. I wonder whether this diversification (presumably inviting some dilution) is really ideal for such a specialized area. However, being less involved for the past 5 years, I should resist my temptation in passing comments in any way.

**Ghosh:** You were the third recipient of the prestigious Senior Noether Award, after Erich Lehmann and Bob Hogg. This bears a strong testimony to your many contributions to nonparametric statistics. Do you distinguish between the classical and modern nonparametric statistics, such as spline smoothing and density estimation, etc.?

**Sen:** I was pleasantly surprised in being awarded, more so as Erich and Bob were in their late seventies or early eighties whereas I was some 15 years younger. However, I am happy to see that, with the exception of Manny Parzen (2005), the awardees after me were all in my age group. Like any other field, in nonparametrics too, the "quick and dirty methods" (Mosteller, 1948), during the 1950s to 1990s, led to an evolution of novel methodology, encompassing a much wider area, including multivariate, sequential and general linear models, along with applications in dosimetry, bioequivalence and clinical trials (survival analysis). In my judgment, a significant annexation to this arena is semiparametrics, sparked in the 1970s by the seminal works of David Cox (1972), and the CART methodology developed in the West Coast a few years later. Spline smoothing, density estimation and, more generally, nonparametric regression are very much in the inner core of nonparametrics and still reside there. The other two developments are somewhat different. Their genesis is in nonparametrics, and yet semiparametrics dominate their nurture. Having said that, I still regard them to be within the core of nonparametrics (which, by no means, is limited to rank statistics or exact distribution-free methods). However, their dependence on computationally intensive tools often calls for data mining or statistical learning tools. There appears to be an undercurrent for these areas to branch out of nonparametrics into separate subdisciplines. I would simply prescribe a single phrase—*Beyond Parametrics*—to include all these evolutions under a common umbrella.

**Schell:** There are four statisticians who have co-authored at least 10 papers with you: Drs. Saleh, Malay Ghosh, Madan Puri and Jana Jurećková. Please talk about these key collaborations.

**Sen:** I prefer to judge the impact of collaboration not by the number of co-authored papers but by their quality. In this respect, the most important one was with Shoutir Chatterjee, which really laid down the foundation of multivariate nonparametrics that was followed through by other subsequent collaborators. I am sure that if Chatterjee were in the USA, we would have had many more joint publications, not only in multivariate nonparametrics but also in time-sequential methods where his impact has been tremendous. Madan Puri and I were both young and full of energy in the late 1960s and it has been a very helpful experience for me to work with him for about 15 years. The other significant impact in the area of sequential nonparametrics (even going over to the Pitman closeness and empirical Bayes methods) has been due to Malay. From Jana, I gained much insight into robust methods. Dr. Saleh's case is somewhat different. He had some ideas on preliminary test estimators in the 1970s. I told him that the asymptotics that had been recently developed could be successfully incorporated in this area. We followed through on these ideas during the next 15 years or so, resulting in a number of publications. However, it became clear to us that such estimators were uniformly dominated by the Stein-type shrinkage estimators, as shown in my 1986 *Sankhya* paper (Sen, 1986). Thus, our work shifted to hierarchical and empirical Bayes analogues of preliminary test estimators. Malay, you may recall our joint paper with you in that arena also.

I should also mention that Manish Bhattacharjee and Yogendra Chaubey have collaborated with me on some interesting ideas in reliability theory and functional estimation.

**Schell:** To date, you have supervised 80 graduate students. Many of them have become successful in later years, either in academia or government. How did you stimulate their enthusiasm for statistics?

**Sen:** A man is known by the company he keeps, and a professor's company is his/her advisees and colleagues. On both counts, I am fortunate. The tradition of outstanding graduate students in Statistics was and is quite strong. In 1969 I started also supervising students in Biostatistics; many of them have good methodological background but were application-oriented. I am happy that in such cases, I could direct them in the right perspectives, albeit they excelled on their own merits. It was statistically expected that some very bright students were in greater need for more advising for concentration and localization of dissertation work. It was my pleasure to see that most of them made it with honor and dignity.

### REFLECTIONS ON STATISTICS, NONPARAMETRICS AND RELIGION

**Schell:** What is your assessment about the future of statistics and biostatistics? Where do you think that we are heading?

**Sen:** I wish I could mutter: Que Sera, Sera. Whatever will be, will be. The future's not ours to (fore)see. Each discipline is going through evolutionary changes, and ours is no exception. Application-oriented and dominated research is reshaping basic research in sciences, while the computer and information technology is totally engulfing the perspectives of the experimental sciences. Statistical science, including statistics and biostatistics, has its genesis in mathematics but has evolved into interdisciplinary fields. Medical studies, clinical trials, environmental health sciences, reliability studies, and genomics and bioinformatics are knocking at the doors. It's a basic task to establish the statistical basis in such experimental fields and develop methodological research to suit the purpose well. Data mining and other computational algorithms are working out well, and yet there is a profound need for implementation of valid, robust and efficient statistical reasoning in such applications. Mathematical abstractions in statistics are fading away, giving way to graphical displays and magical numerical outputs from the supercomputers. Alas, I wish I could attach some validatory statistical interpretations to them. The Bayesian methods are promising, albeit they need to be tuned with proper priors, not the vague ones. Having said so, I am also very hopeful that smart methodological researchers in statistics will find a way out to salvage their methodological works before unconditionally submitting to the wind pipes of data mining.

**Schell:** Do you think that nonparametric methods are appreciated sufficiently by the statistics community today?

**Sen:** Is there any way not to do that? The limitations of the classical parametrics are becoming so evident that either the choice is to surrender to data mining or to go "beyond parametrics." On a positive side, nonparametrics have a natural appeal regarding their scope of applicability and model flexibility. On the negative side, they may inherit larger sample size requirements in order to have adequate precision of drawn statistical conclusions. Fortunately, with the advent of modern computers, large data sets now dominate statistical modeling and analysis, and hence, nonparametrics are being received increasingly by the statistics community. Of course, I must warn that use of data mining tools may not automatically qualify for inclusion in this prescription, and hence, that needs to be appraised properly.

**Ghosh:** One of your favorite pastimes is studying books on religion, especially those written on Ramakrishna, Vivekananda or Saradamani. You also write poetry in both English and Bengali. Do you want to comment on that?

**Sen:** I confess that I do not qualify for tennis or contract bridge nor have any talent for piano or violin. Moreover, during my school days, I had little appetite for literature or newspaper reading. My mother used to worry about my prospects. Just three months after the matriculation examination, I started reading classical novels of contemporary Bengali writers, but did not get much out of it. In the midst of that, my reading the "Sanchayita" of Rabindra Nath Tagore completely changed my views. The more I read the more I was fascinated by the lyrical powers of this great poet of all times. The collection on "Puja" and "Prakriti" were the cream of his understanding of the nature and the lord in a broad sense. Years later, I looked first into the writings of Swami Vivekanada which took me to another world. In due course I also read the "Gospel of Ramakrishna" with deep curiosity. I started wondering how they realized God in their own way, and yet how parallel were their lines of thinking. I am religious but am neither an orthodox nor a fundamentalist. My feeling is almost the way Tagore depicted in his "Geetanjali":

> *Offer thy heart, mind and soul at the feet of the almighty, without any expectation; Suddenly you would realize how insignificant we are in this time and scale of perception.*

Tagore's approach was a lyrical one which suppresses all pains of body and mind and enables one to concentrate in appreciating the superperson in our life-cycle. The origin of poetry lies in the same appreciation. Many things that you cannot express in plain prose can be composed succinctly in a verse, conveying the deeper meaning which reaches the heart easily. To me this is the ideal way of realization of whatever I may aspire. My composition of little poems in both Bengali and English is far from being perfect or on par with contemporary works by others. Yet that gives me a sense of esteem that I love to have. I hope to be able to dip into this in the near future.

**Ghosh:** I understand that you are planning to retire officially in a few years. If I know you, you are never going to give up statistics. But do you have any other plans after retirement?

**Sen:** I would like to take off from in-class teaching and supervision of doctoral students, culminating in an official retirement in the near future. However, that should not put a roadblock to my pursuit of statistical reasoning in a greater domain. I am fascinated by the current development of bioinformatics and environmetrics; there is a tremendous scope for statistical reasoning in such contexts. Quality of life is another area that attracts me more at this stage. The religious inclinations I have should help me in appraising this aspect in a broader spectrum. But, having said so, I must also say that retirement is a natural phase and I should accept it in a natural way mingled with the present state of my mind.

**Ghosh:** Is there anything else that you wish to talk about that we failed to ask you?

**Sen:** Malay and Michael, I am indeed very happy that you have undertaken this conversation with me. I would like to express my thanks to the UNC School of Public Health Dean Professor Barbara Rimer and Ms. Martha Monnett for their interest and arranging this conversation taking place at the Speech Communication Center, UNC.

Let me take this opportunity to express my appreciation for all my colleagues and others in a few lines (adapted from Tagore):

> *Whoever imparted splashes of happiness in my life, I bow to you, and whoever inflicted sufferings in my heart, I bow to you too. All who have lighted the candle of love in my life, showing the way to appreciate everyone, I bow to you too. Whatever came in my way, touched my heart, or gone far away, albeit, in a distant path, I bow to you all in the same way. Knowingly or not, admittedly or not, I have realized Thou: Oh, Holy Mother, and bowing in prostration, I long for you. Statistics is indispensable in every walk of life and science; at this bend of the walk, let hope and faith guide my conscience.*